\documentclass[11pt]{article}
\usepackage{amsmath,amssymb,color,graphics,epsfig}

\usepackage{amsfonts}

\newcommand{\hoch}[1]{$\, ^{#1}$}

\newcommand{\be}{\begin{equation}}
\newcommand{\ee}{\end{equation}}
\newcommand{\bea}{\setlength\arraycolsep{2pt} \begin{eqnarray}}
\newcommand{\eea}{\end{eqnarray}}

\def\ft#1#2{{\textstyle{\frac{\scriptstyle #1}{\scriptstyle #2} } }}
\def\fft#1#2{{\frac{#1}{#2}}}

\def\0{{\sst{(0)}}}
\def\1{{\sst{(1)}}}
\def\2{{\sst{(2)}}}
\def\3{{\sst{(3)}}}
\def\4{{\sst{(4)}}}
\def\5{{\sst{(5)}}}
\def\6{{\sst{(6)}}}
\def\7{{\sst{(7)}}}
\def\8{{\sst{(8)}}}
\def\sst#1{{\scriptscriptstyle #1}}

\begin{document}

\begin{flushright}
\end{flushright}

\vspace{25pt}
\begin{center}
{\large {\bf Charged Rotating AdS Black Hole and Its Thermodynamics\\ in Conformal Gravity}}

\vspace{10pt}
Hai-Shan Liu\hoch{1} and H. L\"u\hoch{2}

\vspace{10pt}

\hoch{1} {\it Institute for Advanced Physics \& Mathematics,\\
Zhejiang University of Technology, Hangzhou 310023, China}

\vspace{10pt}

\hoch{2}{\it Department of Physics, Beijing Normal University,
Beijing 100875, China}

\vspace{40pt}

\underline{ABSTRACT}
\end{center}

We obtain the charged rotating black hole in conformal gravity.  The metric is asymptotic to the (anti-)de Sitter spacetime.  The contribution to the metric from the charges has a slower falloff than that in the Kerr-Newman AdS black hole. We analyse the global structure and obtain all the thermodynamical quantities including the mass, angular momentum, electric/magnetic charges and their thermodynamical conjugates. We verify that the first law of thermodynamics holds.  We also obtain the new neutral rotating black holes that are beyond Einstein metrics.  In contrast to the static ones, these rotating black holes have no parameters associated with the massive spin-2 hair.

\vfill {\footnotesize Emails: hsliu.zju@gmail.com\ \ \ mrhonglu@gmail.com}

\thispagestyle{empty}

\pagebreak



\newpage

\section{Introduction}

The construction of rotating black holes is much subtler than that of static ones. Inspired by string theory, the Kerr metric \cite{kerr} was generalized to arbitrary dimensions \cite{myersperry}. Motivated by the AdS/CFT correspondence, the four-dimensional Kerr-AdS black hole \cite{Carter:1968ks} has been generalized to five \cite{Hawking:1998kw} and arbitrary dimensions \cite{glpp1,glpp2}, even with general NUT parameters \cite{Chen:2006xh}. Adding charges to these neutral rotating black holes is a difficult task.  In fact, aside from the Kerr-Newman solution \cite{newman,Carter:1968ks}, there is no known charged rotating black hole in Einstein-Maxwell theory beyond four dimensions, with or without a cosmological constant.  The situation improves in supergravities, where enhanced global symmetry can be used to generate charged solutions from the static Schwarzschild or rotating Kerr metrics.  For example, the three-charge rotating black hole in five dimensional $U(1)^3$ supergravity was constructed by using the solution generating technique \cite{cveticyoum}.  However, in gauged supergravities, such global symmetries are broken by the scalar potentials that yield the anti-de Sitter (AdS) vacua.  There is no known solution generating technique and solutions are much harder to come by. There have been considerable efforts in constructing charged rotating AdS black holes in gauged supergravities, yielding, for example, the general rotating black holes in five dimensional Einstein-Maxwell gauged supergravity \cite{cclp} and $U(1)^3$ gauged supergravity \cite{wu}. One motivation of constructing these rotating solutions is that the static charged black holes in gauged supergravities suffer from having naked singularities in the supersymmetric limit and the resolution requires rotations \cite{susybh1,susybh2}.

Charged black holes can also emerge in conformal gravity in four dimensions and the spherically-symmetric solution was obtained in \cite{Riegert:1984zz}.  Conformal pure gravity is constructed by the Weyl-squared term.  There have been considerable interests in this theory for decades.  Recently, it was shown that the Einstein-Weyl gravity exhibits critical phenomenon at some special point of the parameter space \cite{lpcritical}, generalizing the result in three dimensions \cite{Li:2008dq}. It was shown that cosmological Einstein gravity can emerge from conformal gravity in the infrared region with the cosmological constant specified by the coupling of the Weyl-squared term \cite{maldconf}. Interestingly the conformal symmetry is preserved when it couples minimally to the Maxwell field.  The Lagrangian is given by
\begin{equation}
e^{-1}{\cal L} =  \ft12\alpha C^{\mu\nu\rho\sigma}
C_{\mu\nu\rho\sigma} + \ft13\alpha F^2\,,\label{conflag}
\end{equation}
where $e=\sqrt{-g}$, $F=dA$ and $C_{\mu\nu\rho\sigma}$ is the Weyl tensor. It was shown in \cite{Lu:2012ag,Li:2012gh} that charged static black holes of (\ref{conflag}) with flat topology provide a useful gravitational background to study strongly-coupled fermionic system where the analytical Green's function $G(\omega,k)$ could be obtained.

Note that the absolute value of the coupling of the Maxwell field can be arbitrary defined by a constant scaling of $A$.  For the purpose of this paper, the sign of the kinetic term is also not essential.  However, since the ghost modes in the gravity sector cannot be avoided, it can be argued that we should choose the standard Lagrangian for the Maxwell field
\begin{equation}
e^{-1}{\cal L}_A=-\ft14 F^2\,,
\end{equation}
so that the ghost modes are restricted in the gravity sector only.  Since it is a trivial step to scale the vector, we shall continue to use the Lagrangian (\ref{conflag}) so that the parameter $\alpha$ does not appear in the local solution.  The equations of motion are
\begin{equation}
\nabla^{\mu} F_{\mu\nu}=0\,,\qquad
-\alpha (2\nabla^\rho\nabla^\sigma +
R^{\rho\sigma})C_{\mu\rho\sigma\nu} - \ft23\alpha (F_{\mu\nu}^2 -
\ft14 F^2 g_{\mu\nu})=0\,.
\end{equation}

It is worth pointing out that conformal gravity can be supresymmetrized in the off-shell formalism.  In ${\cal N}=1$, $D=4$ off-shell supergravity, the bosonic fields consist of the metric, a vector and a complex scalar $S + {\rm i} P$. The fermionic field involves only the off-shell gravitino $\psi_\mu$.  Up to and including the quadratic order in curvature, the theory allows four independent super-invariants. These comprise a ``cosmological term \cite{lpsw},'' the Einstein-Hilbert term \cite{sw,fv}, and two quadratic-curvature terms \cite{ledu}, one formed using the square of the Weyl tensor, and the other formed using the square of the Ricci scalar.  These higher-derivative off-shell supergravities offer a wealthy number of supersymmetric vacua, including AdS, Lifshitz, Schr\"odinger and gyratons \cite{luwangsusylif,liulu,lpgyr,llpp}. The Lagrangian (\ref{conflag}) turns out to be the bosonic part of a super invariant, except that an analytical continuation of $A\rightarrow {\rm i}\, A$ was performed.

In section 2, we give the local solution of charged rotating AdS black hole. It contains four non-trivial parameters associated with the mass, angular momentum and electric/magnetic charges.  We also obtain a new neutral rotating black hole that is not an Einstein metric, but conformal to the Kerr-AdS metric.

In section 3, we analyse the global structure.  We obtain all the conserved quantities and their thermodynamic conjugate pairs.  Owing to the metric complexity from the rotation and higher-derivative nature of the theory, we give some detail in evaluating the temperature, entropy, mass and angular momentum.  The calculation for mass and momentum is rather subtle and the usual methods in the market typically give divergent answers.  We adopt a Noether charge method and obtain the result that is consistent with the first law of thermodynamics.

    In section 4, we consider the extremal limit of the black holes, and obtain
the near-horizon geometry of the extremal solution.  In section 5, we examine the local Plebanski-type solution.  The paper is concluded in section 6.

\section{Rotating (A)dS black hole}

\subsection{Dyonic solution}

In four dimensions, a black hole can carry both electric and magnetic charges. We find that conformal gravity admits the following dyonic rotating black hole:
\begin{eqnarray}
ds^2&=&\rho^2\Big(\fft{dr^2}{\Delta_r} + \fft{d\theta^2}{\Delta_\theta}\Big) + \fft{\Delta_\theta\sin^2\theta }{\rho^2} \Big(a dt - (r^2 + a^2)\fft{d\phi}{\Xi}\Big)^2 - \fft{\Delta_r}{\rho^2} \Big(dt - a \sin^2\theta\fft{d\phi}{\Xi} \Big)^2 \,, \cr
A&=&\fft{q\,r}{\rho^2} \Big(dt - a \sin^2\theta \fft{d\phi}{\Xi} \Big) + \fft{p\,\cos\theta}{\rho^2} \Big(a dt - (r^2 + a^2)\fft{d\phi}{\Xi} \Big) \,,
\label{chargedrot}
\end{eqnarray}
where,
\begin{eqnarray}
&&\rho^2 = r^2 + a^2 \cos^2\theta \,, \qquad \Delta_\theta = 1 +\ft13\Lambda a^2 \cos^2\theta \,, \qquad \Xi = 1+\ft13\Lambda a^2\,,\cr
&&
\Delta_r = (r^2 + a^2)(1 - \ft13\Lambda r^2) -2m r + \fft{(p^2 + q^2) r^3}{6m}\,.
\label{delta}
\end{eqnarray}
If we turn off the charge parameters $(p,q)$, the solution becomes the known Kerr-AdS$_4$ black hole, which is Einstein. If instead we let the rotating parameter $a$ be zero, it becomes a static charged black hole, but with one less parameter than the general spherically-symmetric solution constructed in \cite{Riegert:1984zz}.  The missing parameter is associated with the massive spin-2 hair.  Note that in the usual Kerr-Newman AdS solution, the electric and magnetic charge parameters $(p,q)$ enter $\Delta_r$ simply as $p^2 + q^2$, rather than the last term of $\Delta$ in (\ref{delta}).  The different falloffs indicate that the Kerr-Newman solution and also the Reissner-Nordstr\o m black hole cannot be embedded in conformal gravity. In four dimensions, in addition to the Kerr-Newman AdS black hole, various charged rotating black holes in gauged supergravities have been constructed \cite{Chong:2004na,Chow:2010sf,Chow:2010fw,Wu:2011zzh}.

The local solution is a special limit of the more general Plebanski-type solution \cite{Mannheim:1990ya} which we shall discuss in section 5.

\subsection{New neutral solution}

As we have mentioned, turning off the dyonic charge $p$ and $q$, the charged rotating black hole becomes the Kerr-AdS solution which is an Einstein metric. It is of interest to obtain new rotating black holes in conformal gravity that is beyond Einstein. To do this, we can start with a more general solution
\begin{eqnarray}
&&\rho^2 = r^2 + a^2 \cos^2\theta \,, \qquad \Delta_\theta = 1 + b \cos\theta + \ft13\Lambda a^2 \cos^2\theta \,, \qquad \Xi = 1+\ft13\Lambda a^2\,,\cr
&&
\Delta_r = (r^2 + a^2)(1 - \ft13\Lambda r^2) -2m r + \fft{(3b^2 + p^2 + q^2) r^3}{6m}\,.
\label{delta0}
\end{eqnarray}
This solution suffers from a conical singularity of the principle orbits for non-vanishing parameter $b$.  Instead of the $S^2$, the level surface becomes the tear-drop with either the south or the north pole being conical. Turning off both $p$ and $q$, we obtain a neutral solution with $m$, $a$ and $b$ parameters.  In order to avoid the conical singularity, we can redefine the parameter $b^2\rightarrow 4 m (-\mu)$ and then send $m$ to zero.  The resulting solution becomes
\begin{eqnarray}
ds^2&=&\rho^2\Big(\fft{dr^2}{\Delta_r} + \fft{d\theta^2}{\Delta_\theta}\Big) + \fft{\Delta_\theta\sin^2\theta }{\rho^2} \Big(a dt - (r^2 + a^2)\fft{d\phi}{\Xi}\Big)^2 - \fft{\Delta_r}{\rho^2} \Big(dt - a \sin^2\theta\fft{d\phi}{\Xi} \Big)^2 \,, \cr
\rho^2 &=& r^2 + a^2 \cos^2\theta \,, \qquad \Delta_\theta = 1 + \ft13\Lambda a^2 \cos^2\theta \,, \qquad \Xi = 1+\ft13\Lambda a^2\,,\cr
\Delta_r &=& (r^2 + a^2)(1 - \ft13\Lambda r^2) -2\mu r^3\,.
\label{delta1}
\end{eqnarray}
Although this metric can be obtained directly from (\ref{chargedrot}) by letting $p^2 + q^2 \rightarrow 12 m (-\mu)$, the point is that in conformal pure gravity, the Kerr-AdS metric and the new solution (\ref{delta1}) are mutually exclusive within our ansatz. In fact if we let $\Delta_r$ be a generic function of $r$ and substitute the ansatz to the equations of motion with vanishing $A$, we find that there are two bifurcated solutions. One is the known Kerr-AdS black hole and the other is (\ref{delta1}).  This is different in the static case, where the corresponding $\Delta_r$ is given by \cite{Riegert:1984zz}
\begin{equation}
\fft{\Delta_r}{r^2} = \ell^{-2} r^2  - \fft{c_0^2-1}{2m} r + c_0 - \fft{2m}{r}\,.
\label{static}
\end{equation}
In this case, both the Schwarzschild and new black holes are contained in the more general solution.  The Schwarzschild black hole is given by $c_0=1$ whilst the new solution is given by letting $c_0\rightarrow 1$ and keeping $(c_0^2-1)/(2m)$ as a constant.  The thermodynamics of the black hole (\ref{static}) involves a massive spin-2 hair \cite{Lu:2012xu}.

\subsection{Conformal to Einstein metrics}

It is of interest to investigate whether the neutral solution is conformal to some Einstein metric.  A natural candidate is the Kerr-AdS metric. It was shown in \cite{Gover:2004ar} that if a metric $ds^2$ is conformal to an Einstein metric $d\tilde s^2$ with $ds^2=\Omega^2 d\tilde s^2$, then one must have
\begin{equation}
\nabla_\mu C^{\mu\nu\rho\sigma} - (D-3) V_\mu C^{\mu\nu\rho\sigma}=0\,,\label{conformalcon}
\end{equation}
for some appropriate $V_\mu$.  The conformal factor can then be obtained by $V_\mu =\partial_\mu \log(\Omega)$.  For the new rotating black hole, we find that $\Omega^2=r^2 \cos^2\theta/\tilde\ell^2$.  Indeed the metric
\begin{equation}
d\tilde s^2 = \fft{\tilde \ell^2}{r^2\cos^2\theta} ds^2\label{delta2}
\end{equation}
with $ds^2$ given in (\ref{delta1}) is Einstein with $\Lambda= -3/\tilde \ell^2$.  In fact the metric (\ref{delta2}) is a local metric of the Kerr-AdS black hole.  If we make a coordinate transformation
\begin{equation}
r\rightarrow \fft{1}{r}\,,\qquad
\cos\theta\rightarrow \fft{1}{\cos\theta}\,,\label{insideout}
\end{equation}
the metric (\ref{chargedrot}) with $p=0=q$ becomes (\ref{delta2}) after appropriate renaming of the parameters.  The general solution (\ref{chargedrot}) is locally conformal invariant under the transformation
(\ref{insideout}) together with appropriate redefinition of parameters. It should be emphasized that although (\ref{delta1}) is locally conformal to Kerr-AdS solution, its global structure is different; it turns the ``inside'' of the Kerr-AdS black hole out.

   It should be pointed out that the condition (\ref{conformalcon}) is
necessary but not sufficient.  The static charged solution (\ref{static}) constructed in \cite{Riegert:1984zz} satisfies (\ref{conformalcon}) but it is easy to verify that the metric is not conformal to Einstein.

\subsection{Kerr-Shield form}

The solutions presented so far are in the Boyer-Lindquist coordinates. They can be cast in the Kerr-Shield form, in which the contributions
of the mass parameter $m$ and the charges enter the metric in the perturbative  fashion.  The Kerr-Shield form of the solution is given by
\begin{eqnarray}
ds^2&=& d\bar s^2 + U k^2\,,\qquad U=\fft{-2m r + \ft{p^2+q^2}{6m} r^3}{\rho^2}\,,\cr
A&=&\fft{q\,r}{\rho^2} k + \fft{p\,\cos\theta}{\rho^2} \Big(a dt - (r^2 + a^2)\fft{d\phi}{\Xi} \Big) \,,\cr
k&=&k_\mu dx^\mu= dt - a\sin^2\theta \fft{d\phi}{\Xi} - \fft{\rho^2 dr}{\Delta_r^0}\,,
\end{eqnarray}
where $\Delta_r^0 =(r^2 + a^2) (1-\fft13 \Lambda r^2)$. The metric $d\bar s^2$ is AdS ; it is given in (\ref{chargedrot}) with $p=0=q$ and $m=0$.  In fact, it is possible to write the solution in the doubled Kerr-Shield form in which the cosmological constant enters the metric linearly as well, as in the case of the general Kerr-AdS-NUT solutions \cite{Chen:2007fs}. It is clear that the similar Kerr-Shield form exists for the neutral rotating black hole.

\section{Thermodynamics}

In this section, we obtain the conserved quantities of the rotating black holes constructed in the previous section. We shall focus on the solutions with negative cosmological constant and set
\begin{equation}
\Lambda=-\fft{3}{\ell^2}\,.
\end{equation}
We shall focus on the detail derivation for the charged solution (\ref{chargedrot}) and simply give the answers for the simpler neutral one (\ref{delta1}) in the last subsection.

\subsection{AdS$_4$ vacuum in the rotating frame}

In order to study the global structure and the thermodynamics of the black hole, we first examine the vacuum solution, which is given by
setting $p=0=q$ and $m=0$.  The parameter $a$, however, does not have to vanish.  The metric is then given by
\begin{eqnarray}
ds^2&=&\fft{\rho^2 dr^2}{(r^2 + a^2)(1+\ell^{-2} r^2)} + \fft{\rho^2 d\theta^2}{1-\ell^{-2}a^2 \cos^2\theta}\cr
&&+
\fft{(r^2 + a^2)\sin^2\theta}{\Xi} (d\phi + a \ell^{-2} dt)^2 -
\fft{(1-\ell^{-2}a^2\cos^2\theta) (1 + \ell^{-2} r^2)}{\Xi} dt^2\,.\label{adsinrot}
\end{eqnarray}
This somewhat complicated metric is in fact the AdS$_4$ in the rotating frame.  To see this, we make the following coordinate transformation from $(r,\theta)$ to $(\rho,\tilde\theta)$:
\begin{equation}
\fft{(r^2 + a^2)\sin^2\theta}{\Xi} = \rho^2 \sin^2\tilde\theta\,,\qquad
\fft{(1-\ell^{-2} a^2\cos^2\theta)(1 + \ell^{-2} r^2)}{\Xi} = 1 + \ell^{-2}\rho^2\,.\label{coortrans}
\end{equation}
We find that the metric (\ref{adsinrot}) becomes the standard AdS$_4$ in global coordinates, but with non-vanishing angular velocity:
\begin{equation}
ds^2 = (1 + \ell^{-2} \rho^2) dt^2 + \fft{\rho^2}{1 + \ell^{-2} \rho^2} +
\rho^2 \Big(d\tilde\theta^2 + \sin^2\tilde \theta\, (d\phi + \ell^{-2} a dt)^2\Big)\,.\label{asymrot}
\end{equation}
Interestingly, as long as $\ell^{-2} a^2 <1$, the ranges of the coordinates $(r,\theta)$ are identical to those of $(\rho,\tilde\theta)$. Since the coordinate transformation (\ref{coortrans}) does not involve in $t$ and $\phi$, this shows that the time coordinate $t$ in the metric (\ref{chargedrot}) is properly defined from the point of view of asymptotic infinity.  Furthermore, the longitudinal coordinate $\phi$ must have a period $2\pi$. Thus we see that the angular velocity at infinity is
\begin{equation}
\Omega_\infty = -\ell^{-2} a\,.
\end{equation}

\subsection{Temperature and angular velocity}

We are now in the position to study global structure.  There are two degenerate Euclidean cycles at $\theta=0$ and $\theta =\pi$.  In both cases, the null Killing vector at the degenerate cycle is given by $\partial/\partial\phi$.  It is easy to verify that the corresponding Euclidean surface gravity is unit, and hence the coordinate $\phi$ can be globally defined to have $2\pi$ period, consistent with the analysis at infinity.

    The horizon of the black hole is located at $r=r_+$, where $r_+$ is the
largest root of the the function $\Delta_r$.  Since $\Delta_r$ is a quartic polynomial in $r$, it is advantageous to express some other variable in terms of $r_+$.  Let us set $p=\sqrt{m}\, \tilde p$ and $q=\sqrt{m}\, \tilde q$, we the find
\begin{equation}
m=\ft12 r_+^{-1}(r_+^2 + a^2)(1 + \ell^{-2} r_+^2) + \ft1{12} (\tilde p^2 + \tilde q^2) r_+^2\,.
\end{equation}
The null Killing vector on the horizon is
\begin{equation}
K = \fft{\partial}{\partial t} + \Omega_H \fft{\partial}{\partial\phi}\,,\qquad \Omega_H=\fft{a(1-\ell^{-2}a^2)}{r_+^2 + a^2}\,.\label{nullkilling}
\end{equation}
The surface gravity on the horizon can be obtained by
\begin{equation}
\kappa^2 =-\fft{g^{\mu\nu}\nabla_\mu K^2 \nabla_\nu K^2}{4K^2}\,.
\end{equation}
We find that
\begin{equation}
\kappa = \fft{3(1 + 3 \ell^{-2} r_+^2)r_+^2 -3a^2(1-\ell^{-2} r_0^2)+ r_+^3(\tilde p^2 + \tilde q^2)}{6r_+(r_+^2 + a^2)}\,.
\end{equation}
Thus the temperature is given by
\begin{equation}
T=\fft{\kappa}{2\pi}\,.
\end{equation}
Note that for the properly defined time coordinate $t$, the scale of the Killing vector is set by $K^0=1$. The angular velocity $\Omega_H$ on the horizon can be read off from the null Killing vector (\ref{nullkilling}).  What enters the first-law of thermodynamics is the difference between the angular velocity on the horizon and that at asymptotic infinity \cite{Caldarelli:1999xj,Gibbons:2004ai}:
\begin{equation}
\Omega=\Omega_H - \Omega_{\infty} = \fft{a(1+ \ell^{-2} r_+^2)}{r_+^2 + a^2}\,.
\end{equation}

\subsection{Entropy}

The entropy of black holes in higher-derivative gravities can be computed by the Wald formula \cite{Wald:1993nt}
\begin{equation}
S=-\ft{1}{8} \int_{r=r_+} \sqrt{h} d^2x\, \epsilon_{ab}\epsilon_{cd}\fft{\partial L}{\partial R_{abcd}}\,.\label{waldform}
\end{equation}
For conformal gravity, we have
\begin{equation}
L=\ft12\alpha C^2 = \ft12 \alpha (R^{\mu\nu\rho\sigma}R_{\mu\nu\rho\sigma} - 2 R^{\mu\nu} R_{\mu\nu} + \ft13 R^2)\,.
\end{equation}
Note that here $\sqrt{h} d^2x$ is the volume element of the horizon, and the $\epsilon_{ab}$ is the antisymmetric tensor in the ${\mathbb R}^2$ direction orthogonal to the horizon.  A typical black hole in higher derivative gravity known in the literature is static, and the application of the Wald formula is straightforward. The computation of the entropy is subtler in rotating case and hence we give some detail. It is convenient to write the metric in the following form
\begin{equation}
ds^2 = g_{rr} dr^2 + g_{\theta\theta} d\theta^2 + g_{\phi\phi} (d\phi + B dt)^2 - C dt^2\,,
\end{equation}
which leads to a natural choice for vielbein:
\begin{equation}
e^{\bar 0} = \sqrt{C} dt\,,\qquad e^{\bar 1}= \sqrt{g_{rr}} dr\,,\qquad
e^{\bar 2}=\sqrt{g_{\theta\theta}} d\theta\,,\qquad e^{\bar 3}=\sqrt{g_{\phi\phi}} (d\phi + B dt)\,.
\end{equation}
The indices $a$ and $b$ in (\ref{waldform}) take the $(\bar 0, \bar 1)$ directions. Thus we have
\begin{equation}
\epsilon_{ab}\epsilon_{cd}\fft{\partial L}{\partial R_{abcd}}=
\ft12\alpha \Big(8 R^{\bar 0\bar 1\bar 0 \bar 1} -4 (R^{\bar 0\bar 0} - R^{\bar 1\bar 1}) - \ft43 R\Big)\,,
\end{equation}
and
\begin{equation}
d^2x\,\sqrt{h}\Big|_{r\rightarrow r_+}  =d\theta d\phi\, \sqrt{g_{\phi\phi} g_{\theta\theta}}\Big|_{r\rightarrow r_+}=\fft{r_+^2 + a^2}{\Xi} \sin\theta d\theta\, d\phi\,.
\end{equation}
We find that the entropy is
\begin{equation}
S=\fft{2\alpha \pi}{\Xi}\Big( 1 + \ell^{-2} r_+^2 + \ft16r_+ (\tilde q^2 + \tilde p^2)\Big)\,.
\end{equation}
Note that by using the above procedure, the contribution to the entropy from the Gauss-Bonnet term can be shown to be a numerical number $4\pi$, as one would expect.

\subsection{Electric/magnetic charges and their potentials}

Since the Maxwell field is of the usual two-derivative theory,
the electric and magnetic charges can be obtained straightforwardly, given by
\begin{equation}
Q=\fft{\alpha}{12\pi}\int {*F} = -\fft{\alpha q}{3\Xi}\,,\qquad
P=\fft{\alpha}{12\pi}\int F = -\fft{\alpha p}{3\Xi}\,.
\end{equation}
The electric potential is given by
\begin{equation}
\Phi_Q = K^\mu A_\mu = \fft{q r_+}{r_+^2 + a^2}\,,
\end{equation}
where $K_\mu$ is the null Killing vector on the horizon, given in (\ref{nullkilling}). (Note that the electric potential at infinity vanishes in our gauge choice.) By the virtual of the electric and magnetic duality, we deduce that the thermodynamical potential associated with the magnetic charge is
\begin{equation}
\Phi_P = \fft{p r_+}{r_+^2 + a^2}\,.
\end{equation}

\subsection{Free energy}

Since the Weyl tensor converges sufficiently for our solution, we can obtain the free energy from the Euclidean action without needing boundary counter terms:
\begin{eqnarray}
F &=&-\fft{\omega_2\alpha }{16\pi} \int_{r_+}^\infty r^2 dr \Big(\ft12
|{\rm Weyl}|^2 + \ft13 F^2\Big)\cr
&=&
\fft{\alpha (a^2-r_+^2)(1 + \ell^{-2} r_+^2)}{12\Xi r_+ (r_+^2 + a^2)}\Big(6 (1 + \ell^{-2} r_+^2) + r_+ (\tilde q^2 + \tilde p^2)\Big)\,.\label{freeenergy}
\end{eqnarray}
There is a subtlety associated with the $F^2$ contribution to the free energy.  It is clear that the electric flux contributes negatively in the $F^2$ term, whilst the magnetic flux contributes positively.  On the other hand, $p$ and $q$ contribute symmetrically as $p^2 + q^2$ in the Weyl-squared term.  Thus if we would simply substitute the solution in the action, we would have obtained a result that is asymmetric in electric and magnetic charges.  The free energy (\ref{freeenergy}) is obtained instead by first setting $\tilde p=0$ and then replacing $\tilde q^2$ by $\tilde q^2 + \tilde p^2$ in the result, by the virtual of electric/magnetic duality.

\subsection{Mass and angular momentum}

The calculation of the energy and angular momentum can be rather subtle for AdS black holes in higher derivative gravities.  If we set $p=0=q$, and hence the metric is Einstein, there are many procedures \cite{destek,Ashtekar:1984zz,Ashtekar:1999jx,Okuyama:2005fg,Pang:2011cs} to obtain the mass and the result is the same, given by
\begin{equation}
E=\fft{2\alpha m}{\ell^2\Xi^2}\,.\label{massnocharge}
\end{equation}
It is the same mass one would get from Einstein gravity, but multiplied by the coupling of the Weyl-squared term.
However, once the $\Delta_r$ in the solution (\ref{chargedrot}) involves an $r^3$ term, all these procedures give rise to divergent results.  Such $r^3$ term can arise because of the contribution of the massive spin-2 hair \cite{Lu:2012xu}, or in our case, it is turned on by the electric or magnetic charges.

     In conformal gravity, owing to the fact that the Weyl tensor converge
rapidly for our black holes, one can calculate the mass as the Noether charge associated with the translational symmetry along the time-like Killing direction.  For a given Killing vector $\xi$, its converged current in conformal gravity was shown to take the form \cite{Lu:2012xu}
\begin{equation}
{\cal T}^{\mu\nu} = C^{\mu\nu\rho\sigma} \nabla_\rho\xi_\sigma -
2\xi_\sigma \nabla_\rho C^{\mu\nu\rho\sigma}\,.
\end{equation}
In order to obtain the appropriate time-like Killing vector associated with the mass, it is necessary first to make a coordinate transformation
\begin{equation}
\phi\rightarrow \phi -a\ell^{-2}\, t\label{norot}
\end{equation}
so that the metric is not rotating at the asymptotic infinity. (See equation (\ref{asymrot}).)  The time-like Killing vector is given by $\xi_1=\partial/\partial t$.  We find that the mass of the black hole is
\begin{equation}
E=-\fft{\alpha}{8\pi} \int \lim_{r\rightarrow \infty} {\cal T}^{01}(\xi_1) =
\fft{2\alpha}{\ell^2 \Xi^2} \Big(m + \fft{a^2}{12m} (q^2 + p^2)\Big)\,.\label{energy}
\end{equation}
The result reproduces the mass (\ref{massnocharge}) when the charges or $a$ vanish.
Analogously, we can obtain the angular momentum which is the conserved charge associated with the rotational invariance, whose Killing vector is $\xi_2=\partial/\partial \phi$.  We find
\begin{equation}
J=-\fft{\alpha}{8\pi} \int \lim_{r\rightarrow \infty} {\cal T}^{01}(\xi_2) =
\fft{2a \alpha}{\ell^2 \Xi^2} \Big(m + \fft{\ell^2}{12m} (q^2 + p^2)\Big)\,.
\label{angmom}
\end{equation}
Note that if we had not made the coordinate transformation (\ref{norot}) and calculated the energy and angular momentum in the rotating frame, the second term in the big brackets in (\ref{energy}) and (\ref{angmom}) would not have been there.  The results would be the same as those of the neutral Kerr-AdS metric.

\subsection{First law of thermodynamics}

Having obtained all the thermodynamical quantities, we can verify that the first law of thermodynamics holds, namely
\begin{equation}
dE = T dS + \Omega dJ + \Phi_Q dQ + \Phi_P dP + \Theta d\Lambda \,.
\end{equation}
Note that in conformal gravity, the cosmological constant $\Lambda$ is an integration constant, and hence it is natural to include it as a thermodynamical variable.  The corresponding potential is given by
\begin{equation}
\Theta = -\fft{3\alpha (r_0^2 + a^2)}{2\Xi^2 r_0} \Big(1 + \ell^{-2} r_0^2 + \ft16 r_0(\tilde q^2 + \tilde p^2)\Big)\,.
\end{equation}
The free energy and the mass of the black hole are related as follows
\begin{equation}
F=E - T S - \Omega J - \Phi_Q Q - \Phi_P P\,.
\end{equation}
The Smarr formula takes the following form
\begin{equation}
E=2 \Theta \Lambda\,.\label{smarr}
\end{equation}
As mentioned earlier, setting the rotation parameter $a$ to zero leads to the static solution in \cite{Riegert:1984zz} with one less parameter. Comparing with the thermodynamics for the (charged) static black hole \cite{Lu:2012xu,Li:2012gh},  we find that the massive spin-2 hair is absent in the rotating solution.

\subsection{Thermodynamics for the neutral black hole}

Having obtained the thermodynamics for the more complicated charged rotating black hole (\ref{chargedrot}), we shall simply give the results for the new neutral rotating black hole (\ref{delta1}). Assuming that the horizon is located at $r_+$, the parameter $\mu$ can be expressed as
\begin{equation}
\mu=\fft{(r_+^2 + a^2)(1+\ell^{-2} r_+^2)}{2r_+^3}\,.
\end{equation}
The entropy and temperatures are given by
\begin{equation}
T=\fft{r_+^2(\ell^{-2} r_+^2-1)-
a^2 (3 + \ell^{-2} r_+^2)}{4\pi r_+ (r_+^2 + a^2)}\,,\qquad
S=\fft{-2\pi\alpha a^2 (1 + \ell^{-2} r_+^2)}{\Xi r_+^2}\,.
\end{equation}
Note that $\alpha$ should be negative for positive entropy in this neutral black hole. The angular velocity and the momentum are given by
\begin{equation}
\Omega=\fft{a(1+ \ell^{-2} r_+^2)}{r_+^2 + a^2}\,,\qquad J=-\fft{2\alpha a \mu}{\Xi^2}\,.
\end{equation}
The energy and the free energy associated with the Euclidean action are
\begin{equation}
E=-\fft{2\alpha a^2\mu}{\ell^2 \Xi^2}\,,\qquad F=\fft{\alpha a^2 (r_+^2-a^2) (1 + \ell^{-2} r_+^2)^2}{2\Xi r_+^3
(r_+^2 + a^2)}\,.
\end{equation}
The potential conjugate to the cosmological constant $\Lambda=-1/(3\ell^2)$ is
\begin{equation}
\Theta = \fft{3\alpha a^2 (r_+^2 + a^2) (1+ \ell^{-2} r_+^2)}{2\Xi^2 r_+^3}
\,.
\end{equation}
It is straightforward to verify that the following thermodynamic relations hold:
\begin{equation}
dE=T dS + \Omega dJ- \Theta d\Lambda\,,\qquad F=E - T S - \Omega J\,.
\end{equation}
The Smarr formula is the same as (\ref{smarr}).  Note that if we turn off the parameter $a$, all the conserved charges and thermodynamical quantities except for the temperature vanish.  For this reason, the static figuration was called a thermal vacuum in \cite{Lu:2012xu}.

\section{Extremal limit and the near-horizon geometry}

Both the charged and the new neutral rotating black holes have extremal limits, for which the temperature vanishes. Typically, there exists a decoupling limit for an extremal black hole in which the near-horizon geometry is blown up and becomes a solution on its own. We study the near-horizon geometry of the extremal solutions in this section.

\subsection{The charged solution}

The charged rotating black hole (\ref{chargedrot}) becomes extremal when
\begin{equation}
\tilde q^2 + \tilde p^2 = \fft{3}{r_+^3}\Big(a^2 - r_+^2 - \ell^{-2}r_+^2 (3r_+^2 + a^2)\Big)\,.
\end{equation}
The decoupling limit is given as follows.  First let us make a coordinate transformation
\begin{equation}
r=r_+ + \epsilon \rho\,,\qquad
t=\fft{(r_+^2 + a^2)}{\epsilon V} \tilde t \,,\qquad
\phi=\tilde \phi + \fft{(r_+^2 + a^2)\Omega_H}{V \epsilon}\, \tilde t
\end{equation}
where
\begin{equation}
V=\fft{1}{2r_+^2} \Big(3a^2 - r_+^2 +\ell^{-2} r_+^2 (3r_+^2 - a^2)\Big)\,.
\end{equation}
We then send $\epsilon\rightarrow 0$, we find that the solution, after dropping the tilde in $t$ and $\phi$, becomes
\begin{eqnarray}
ds^2 &=& \fft{r_+^2 + a^2\cos^2\theta}{V} \Big(\fft{d\rho^2}{\rho^2} -
\rho^2 dt^2 + \fft{V\,d\theta^2}{\Delta_\theta} \Big)\cr
&&+\fft{\Delta_\theta\sin^2\theta}{r_+^2 + a^2 \cos^2\theta}\Big
(\fft{r_+^2 + a^2}{\Xi} d\phi + \fft{2a r_+}{V} \rho dt\Big)^2\,,\cr
A &=& -\fft{\sqrt{M}}{r_+^2 + a^2 \cos^2\theta} \Big[((r_+^2 - a^2 \cos^2\theta)\tilde q + 2 a r_+ \cos\theta\, \tilde p)\fft{\rho\,dt}{V}\cr
&&\qquad\qquad +\Big( a r_+ \sin^2\theta\, \tilde q + (r_+^2 + a^2)\cos\theta\,\tilde p\Big) \fft{d\phi}{\Xi}\Big]\,.
\end{eqnarray}
Making a further coordinate transformation
\begin{equation}
\rho=r + \sqrt{1 + r^2} \cos\tau\,,\qquad t=\fft{\sqrt{1 + r^2}\,\sin\tau}{r +
\sqrt{1 + r^2}\, \cos\tau}\,,
\end{equation}
with an appropriate shifting of $\phi$, and gauge choice, we find that the near-horizon metric and $A$ become
\begin{eqnarray}
ds^2 &=& \fft{r_+^2 + a^2\cos^2\theta}{V} \Big(\fft{dr^2}{1 + r^2} -
(1 + r^2) d\tau^2 + \fft{V\,d\theta^2}{\Delta_\theta} \Big)\cr
&&+\fft{\Delta_\theta\sin^2\theta}{r_+^2 + a^2 \cos^2\theta}\Big
(\fft{r_+^2 + a^2}{\Xi} d\phi + \fft{2a r_+}{V} r d\tau\Big)^2\,,\cr
A &=& -\fft{\sqrt{M}}{r_+^2 + a^2 \cos^2\theta} \Big[((r_+^2 - a^2 \cos^2\theta)\tilde q + 2 a r_+ \cos\theta\, \tilde p)\fft{r\,dt}{V}\cr
&&\qquad\qquad +\Big( a r_+ \sin^2\theta\, \tilde q + (r_+^2 + a^2)\cos\theta\,\tilde p\Big) \fft{d\phi}{\Xi}\Big]\,.
\end{eqnarray}

\subsection{The neutral solution}

The solution becomes extremal when
\begin{equation}
\mu=\fft{1 + \ell^{-2} r_+^2}{r_+ (3 + \ell^{-2} r_+^2)}\,,\qquad a^2 = \fft{r_+^2 ( \ell^2 r_+^2 -1)}{3 + \ell^{-2} r_+^2}\,.
\end{equation}
Using the same procedure discussed above, we obtain the near-horizon metric of the extremal solution
\begin{eqnarray}
ds^2 = \fft{r_+^2 + a^2\cos^2\theta}{V} \Big(\fft{d\rho^2}{\rho^2} -
\rho^2 dt^2 + \fft{V\,d\theta^2}{\Delta_\theta} \Big)
+\fft{\Delta_\theta\sin^2\theta}{r_+^2 + a^2 \cos^2\theta}\Big
(\fft{r_+^2 + a^2}{\Xi} d\phi + \fft{2a r_+}{V} \rho dt\Big)^2,
\end{eqnarray}
where
\begin{equation}
V=\fft{-3 + 6\ell^{-2} r_+^2 + \ell^{-4} r_+^4}{3 + \ell^{-2} r_+^2}\,.
\end{equation}

For both charged and neutral solutions, the near-horizon geometry is an $S^2$ bundle over AdS$_2$.

\section{General dyonic Kerr-AdS-NUT solution}

In the charged rotating black hole (\ref{chargedrot}), the radial coordinate $r$ and the latitude coordinate $\theta$ are quite different. However, the coordinates $r$ and $a\cos\theta$ enter the metric symmetrically.  Let $x=r$ and $y=a\cos\theta$, the general dyonic Kerr-AdS-NUT solution can be put in the Plebanski form
\begin{eqnarray}
ds^2 &=& - \fft{X}{x^2 + y^2} (dt + y^2 d\psi)^2 + \fft{Y}{x^2 + y^2} (dt - x^2 d\psi)^2 + \fft{x^2 + y^2}{X} dx^2 + \fft{x^2 + y^2}{Y} dy^2\,,\cr
A&=& \fft{q\, x}{x^2 + y^2} (dt + y^2 d\psi) + \fft{p\, y}{x^2 + y^2} (dt - x^2 d\psi)\,,
\end{eqnarray}
where the functions $X$ and $Y$ depend only on $x$ and $y$ respectively and they can be solved by the polynomials
\begin{equation}
X=c_0 + c_1 x + c_2 x^2 + c_3 x^3 - \ft13 \Lambda x^4\,,\qquad
Y=c_0 + d_1 y - c_2 y^2 + d_3 y^3 - \ft13 \Lambda y^4\,,
\end{equation}
subjected to the following constraint
\begin{equation}
3(d_1 d_3 -c_1 c_3) = p^2 + q^2\,.
\end{equation}
The solution was obtained in \cite{Mannheim:1990ya}.  At the first sight, the solution contains seven parameters.  However, the metric has a scaling symmetry $(x,y)\rightarrow \lambda (x,y)$ together with appropriate scalings of $(t,\psi)$ and the integration constants.  One of the seven parameters is discrete taking either $\pm1$ or 0.  The remaining six continuous parameters are associated with the cosmological constant, mass, angular momentum, NUT and electric/magnetic charges. We can use an overall conformal scaling of the metric with the conformal factor $1/(1-xy)^2$ to fix this symmetry, leading to the generalization of the Plebanski-Diemenienski solutuion.

The rotating black holes in section 2 is a special case of the more general Plebanski-type solution.  To see this, we note that for compact horizon geometry, the coordinate $y$ is compact and take the range $(b,a)$.  In order to avoid naked close time-like circle, only one Killing vector $\partial_\psi + \omega \partial_t\equiv\partial_\phi$ is periodic on and outside of the horizon, for an appropriate constant $\omega$.  This implies that we must have $b=-a$ such that only the cycle associate with $\partial_\phi$ becomes degenerate at both the ``north'' and ``south'' poles $y=\pm a$.  We find that this implies that $Y$ contains only even powers of $y$, namely $Y=(y^2-a^2) (c_0 + a^2 \ell^{-2} y^2)$.  It is then straightforward to put the form of the solution in the standard Boyer-Lindquist form (\ref{chargedrot}) in which the NUT parameter is removed.

\section{Conclusions}

In this paper, we obtain the charged and new neutral rotating black holes in conformal gravity in four dimensions.  The charged solution takes the analogous form as the Kerr-Newman-AdS, but with the charge contributing a slower fall-off asymptotically.  The neutral solution and the usual Kerr-AdS metric do not appear to be generalizable to a more general black hole of pure conformal gravity.  In conformal gravity, local solutions are defined up to an arbitrary conformal factor.  However, black holes are specified by not only the local metric, but also the global structure.  Conformal factors that alter neither the asymptotic infinity  nor the horizon give rise to an equivalent class of black holes, for which the thermodynamical properties are the same.

    We obtain all the conserved quantities including the mass, angular momentum
and charges, as well as all the thermodynamical quantities including temperature, entropy, free energy {\it etc.} We verify that the first law of the thermodynamics holds.  Many of these quantities, energy and angular momentum in particular, are not straightforward to calculate in higher-derivative gravities.  Our exact rotating solutions of conformal gravity provide good examples to understand the energy in higher derivative gravity.  Many known methods of calculating the energy for our black holes give rise to divergent result.  A general formula of energy is still lacking.

The existence of the exact non-trivial rotating black hole solution allows us to demonstrate that the black hole thermodynamics hold even in gravity theories plagued with ghost massive spin-2 modes.  It suggests that conformal gravity, coupled with scalars or spinors, can be used to study strongly coupled condensed matter system such as superconductor or non-Fermi liquids.  Our exact solutions also provide useful backgrounds to study the general properties of higher-derivative gravities, which arise naturally in string theory.

\section*{Acknowledgement}

We are grateful to Yi Pang, Chris Pope and Zhao-Long Wang for useful discussions.  The research of L\"u is supported in part by the NSFC grants 11175269 and 11235003.

\end{document}